\documentclass[10pt,letterpaper]{article}
\usepackage[top=0.85in,left=2.75in,footskip=0.75in]{geometry}

\usepackage{amsmath,amssymb}

\usepackage{changepage}

\usepackage[utf8x]{inputenc}

\usepackage{textcomp,marvosym}

\usepackage{cite}

\usepackage{nameref,hyperref}

\usepackage[right]{lineno}

\usepackage{microtype}
\DisableLigatures[f]{encoding = *, family = * }

\usepackage[table]{xcolor}

\usepackage{array}

\newcolumntype{+}{!{\vrule width 2pt}}

\newlength\savedwidth

\raggedright
\usepackage[aboveskip=1pt,labelfont=bf,labelsep=period,justification=raggedright,singlelinecheck=off]{caption}

\makeatletter
\renewcommand{\@biblabel}[1]{\quad#1.}
\makeatother

\usepackage{lastpage,fancyhdr,graphicx}
\usepackage{epstopdf}
\fancyheadoffset[L]{2.25in}
\fancyfootoffset[L]{2.25in}
\begin{document}
\vspace*{0.2in}

\begin{flushleft}
{\Large
\textbf\newline{Dataflow programming for the analysis of molecular dynamics with AViS, an analysis and visualization software application} 
}
\newline
\\
Kai Pua\textsuperscript{1\textcurrency},
Daisuke Yuhara\textsuperscript{1},
Sho Ayuba\textsuperscript{1},
Kenji Yasuoka\textsuperscript{1*}
\\
\bigskip
\textbf{1} Department of Mechanical Engineering, Keio University, 3-14-1 Hiyoshi, Kohoku-ku, Yokohama, Kanagawa, JAPAN 223-8522
\\
\bigskip

\textcurrency Current Address: Department of Computer Science, Graduate School of Information Science and Technology, University of Tokyo,
7-3-1 Hongo, Bunkyo-ku, Tokyo, JAPAN 113-8656

* yasuoka@mech.keio.ac.jp

\end{flushleft}
\section*{Abstract}
The study of molecular dynamics simulations is largely facilitated by analysis and visualization toolsets. However, these toolsets are often designed for specific use cases and those only, while scripting extensions to such toolsets is often exceedingly complicated. To overcome this problem, we designed a software application called AViS which focuses on the extensibility of analysis. By utilizing the dataflow programming (DFP) paradigm, algorithms can be defined by execution graphs, and arbitrary data can be transferred between nodes using visual connectors. Extension nodes can be implemented in either Python, C++, and Fortran, and combined in the same algorithm. AViS offers a comprehensive collection of nodes for sophisticated visualization state modifications, thus greatly simplifying the rules for writing extensions. Input files can also be read from the server automatically, and data is fetched automatically to improve memory usage. In addition, the visualization system of AViS uses physically-based rendering techniques, improving the 3D perception of molecular structures for interactive visualization. By performing two case studies on complex molecular systems, we show that the DFP workflow offers a much higher level of flexibility and extensibility when compared to legacy workflows. The software source code and binaries for Windows, MacOS, and Linux are freely available at \url{https://avis-md.github.io/}.


\section*{Introduction}

The study of molecular behaviour through analysis and visualization of Molecular Dynamics (MD) simulations is becoming more widely adopted with the increasing processing power of modern computers. A large number of software was developed for this purpose, each with a give-and-take balance between analysis and visualization flexibility. For example, analysis-focused libraries such as MDAnalysis \cite{michaud2011mdanalysis} \cite{gowers2016mdanalysis}, MDTraj \cite{mcgibbon2015mdtraj}, and cpptraj \cite{cpptraj} provide the necessary toolsets to analyze complex molecular systems to extract relevant information, while some analysis software such as VMD \cite{humphrey1996vmd}, AVOGADRO \cite{hanwell2012avogadro}, PyMOL \cite{delano2002pymol}, and VESTA \cite{momma2011vesta}, apart from their analysis capabilities, also provide tools to visualize the results. While most analysis software offer very extensive analysis capabilities, and they allow complex modifications and extensions to the framework, a deep knowledge of their respective scripting interfaces are required, and hard constraints applied to valid scripts are a hurdle for novice users, even those well versed in bare-bone analysis programming. Therefore, common analysis methods such as hydrogen bonds or Common Neighbor Analysis are often provided in most applications and new algorithms are seldom implemented for integration with the software, except for advanced users. Other software such as OVITO \cite{stukowski2009visualization} uses a combined visual workflow of analysis and visualization, and enabling complicated techniques to be separated and applied in steps. However, a large dependency on the API is still required when writing custom modules, and the flexibility of data transfer between functions is somewhat limited. In this paper, we seek to improve these limitations by proposing an Analysis and Visualization Software (AViS) application from scratch which utilizes the dataflow programming (DFP) paradigm and allows for graphical designing and debugging of algorithms. This study also demonstrates the flexibility and API simplicity of AViS and DFP in the analysis and visualization of MD data. By using AViS, we are able to simplify and accelerate various analysis and visualization tasks, as described in the sections that follow.

\section*{Methods}

\subsection*{Software Overview}

\begin{figure}[ht!]\centering
    \includegraphics[width=0.6\textwidth]{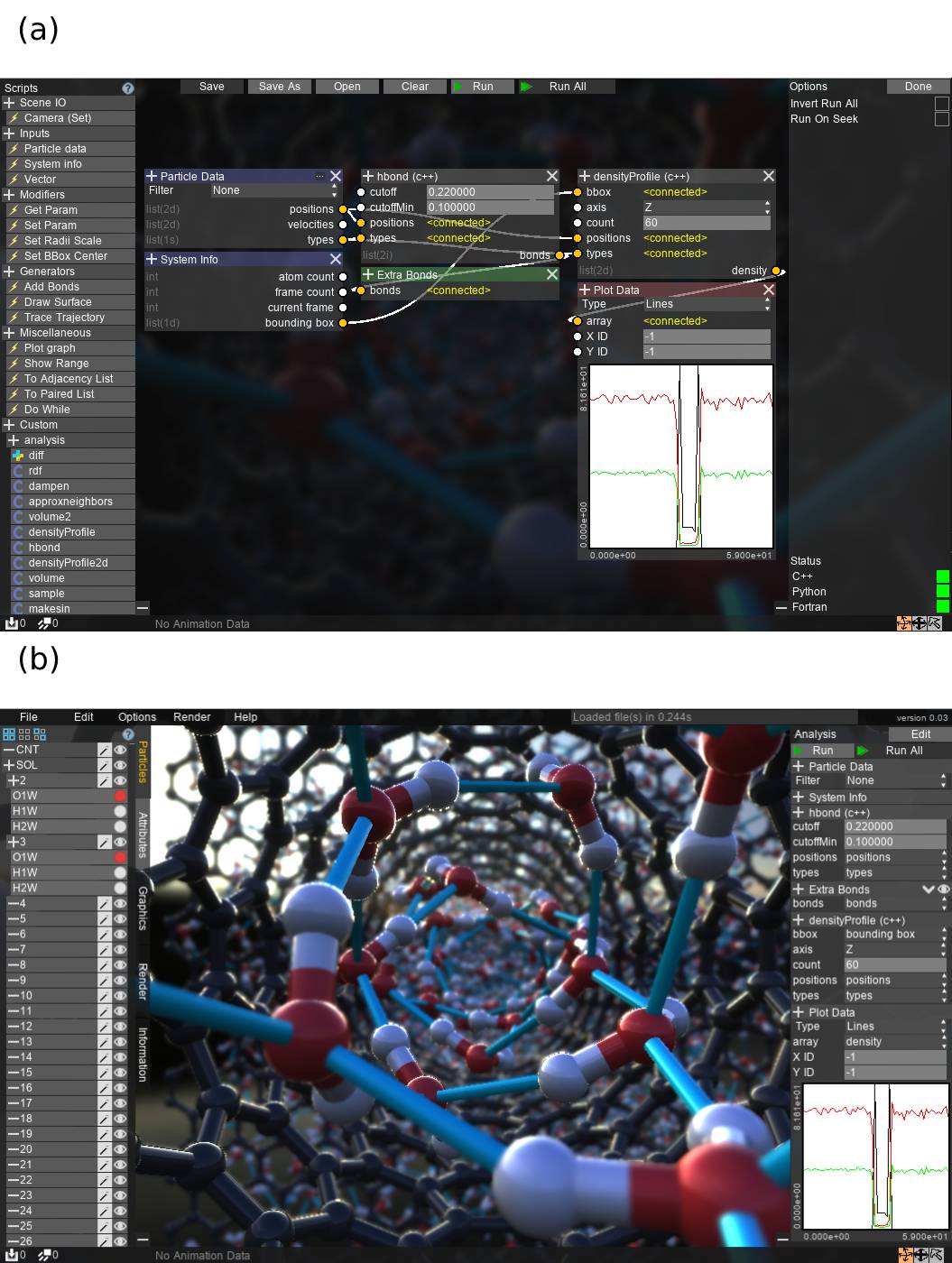}
    \caption{\label{fig:ui} {\bf Application screenshot.} A screenshot of AViS, with its analysis and visualization modes. (a) Analysis mode of AViS. The graphical algorithm to compute the hydrogen bonds and density profile of carbon nanotubes are set up. (b) Visualization mode of AViS. The system as well as the results of the analysis is visualized. Various realtime effects such as Bloom, Ambient Occlusion, and Depth of Field are also applied to the 3D view. }
\end{figure}

AViS is an OpenGL-based application written in C++. The software is portable and runs natively on various Microsoft Windows / macOS / Linux versions. As an integrated software application, AViS provides a unified environment for analysis and visualization. The user interface of AViS is shown in Fig. \ref{fig:ui}.

\subsection*{Analysis}

\subsubsection*{Application of Dataflow Programming}

As existing analysis paradigms lack flexibility, we designed AViS with the DFP paradigm in its analysis framework, implemented as execution graphs. An execution graph contains multiple nodes, which are evaluated in order and their outputs are passed on to other nodes. Due to the black-box nature of DFP, users do not need to control data conversion and copying, and thus a combination of nodes implemented in different languages in the same graph is allowed. Therefore, new nodes can be created by the user using either Fortran, Python, or C++. Each node exposes its input/output variables in the user interface (UI), which can then be connected to any counterpart with a compatible type. In addition to integer and floating-point values, AViS also supports the data transfer of multi-dimensional variable-length arrays. A list of internal nodes provided by AViS is shown in \nameref{s1_table}.

\subsubsection*{Scripting}

Due to the multi-language compatibility, a generic and easy-to-understand format must be enforced for analysis scripts. In this case, we specifically designed the software with no API whatsoever; all relevant code is appended by the internal pre-processor. This makes writing scripts for AViS extremely easy as the user does not have to learn about any application-specific function calls or data structures. As a replacement, we require that all variables that the user wishes to expose to other nodes be preceded with a short comment. An example of valid analysis scripts written in Fortran, Python, and C++ is shown in \nameref{s1_app}, \nameref{s2_app}, and \nameref{s3_app}, respectively, and they can be used together, as shown in \nameref{s1_fig}. Compilers available in the system are used for compilations, with support for gcc, clang, MSBuild and Python 3.7. AViS can still work with other languages even if certain compilers are not available. A summary of comment syntax for each language is shown in \nameref{s4_app} through \nameref{s6_app}.

To allow for interoperability between languages, C++ and Fortran scripts are edited internally before being compiled and loaded as dynamic libraries. Additional code is inserted in place of the comments, so the line numbers in the error messages will still match the original file. For Fortran scripts, subroutines that expose the shape and contents of arrays to AViS are automatically generated and appended. This allows AViS to interact with multi-dimensional variable-length arrays in Fortran scripts, with no effort on the user's part. For Python, all variables are accessed directly using the Python C API, and arrays use the Numpy\cite{numpy} C API.

\subsubsection*{Attributes}

Compared to classic data processing pipelines and single-file analysis workflows, the graph workflow provides a higher level of freedom for data transfer. As data flow is defined solely in the UI, very generic scripts and algorithms may be written, without special consideration of the data source. To ease the analysis of values assigned to each atom, such as order parameter values, AViS also allows for the definition of an arbitrary number of ``attributes'' in the UI. The values can be read from, and written to, at any point in the execution. Operations with attributes are defined in the graph UI, so complicated algorithms involving the reuse of multiple variables can be achieved easily. Attributes are also useful for algorithm debugging purposes as they allow for the visualization of intermediate outputs.

\subsection*{Data Input}

As many different types of simulation software give rise to a variety of file formats, it is vital that AViS supports as many formats as possible. AViS natively supports the importing of configuration and trajectory files from popular formats such as GROMACS \cite{hess2008gromacs}, LAMMPS \cite{plimpton2007lammps}, and Protein Data-Bank \cite{berman2006protein} files. Other formats can also be expanded for AViS by writing a C plugin and putting the compiled library in the designated folder. A generic Space-Separated Values (SSV) format is also available for loading trajectories containing arbitrary attributes without writing a plugin. An example file is shown in \nameref{s7_app}. By importing attributes, users can perform analysis on complex molecules without being constrained by the existing attribute list of the software or the file formats. An example of this usage for a system of liquid crystals as shown in \nameref{s2_fig}, where the orientation vectors are imported as the rotx, roty, and rotz, attributes and may be analyzed and visualized without the need for software extensions. The data import and analysis frameworks are entirely seperated, so different files written in different formats can be easily analysed using the same analysis graph.

\subsection*{Visualization}

As graphics cards become more powerful, visualization programs should exploit the capabilities of the hardware with the newest methods and APIs. AViS uses the OpenGL 3.3 standard, which is the most recent version that is also supported on most computers. All atoms and bonds are drawn as sprites using ray-tracing methods, and all polygon meshes, such as protein strips, are generated on-the-fly using shaders. The lighting model is physically-based and uses pre-processed radiance maps and a Fresnel term \cite{shlick}, improving the 3D perception of molecules. The deferred rendering method is adopted for the rendering path to allow for sophisticated lighting techniques and improved performance of the UI. Details of the rendering procedure are shown in \nameref{s3_fig}. By combining these methods, we are able to achieve interactive to real-time frame-rates on typical computers, even for systems with millions of atoms. As the deferred rendering path saves intermediate results, we further improve performance of the interface by only redrawing the parts of the rendering path that are required. For example, when the user changes the lighting intensity, the geometry buffer is retained and only the lighting buffer is recalculated. The intermediate buffers also allow for sophisticated lighting effects such as Depth-of-Field and Screen-Space Ambient Occlusion.

In the place of a scripting interface, AViS provides a comprehensive collection of operation nodes. These nodes allow arbitrary script inputs/outputs to interact with the loaded data and visualization state. For example, the following non-exhaustive operations can be performed using nodes:

\begin{itemize}
    \item Querying all positions of current atoms that are visible or in a certain region
    \item Modifying the color/radius/visibility of atoms
    \item Creating bonds between atoms
\end{itemize}


\section*{Results and Discussion}

Two case studies are used to demonstrate the flexibility and extensibility of AViS. Here, we look at the formation of Lennard-Jones (LJ) droplets and clathrate hydrates. We choose these systems as their analysis process is complex enough to show the capabilities of AViS. As the total file sizes of the two simulations considered are about 7 and 14 Gigabytes, respectively, the data for both simulations are prepared on a remote file storage system. However, we show that analysis and visualization can still be performed on them by utilizing the remote file handling capability of AViS. All scripts used in these studies are provided as supplementary information, implemented in C++. Some scripts are also translated to Python and Fortran for reference. The analysis results remain consistent when replacing nodes with one generated from a different language. We encourage the reader to refer to the comments in each script detailing the implementation for each analysis step.

\subsection*{Analysis Flexibility}

For the first case study, we performed the tracking of nucleated clusters in a LJ droplet system \cite{ayuba}. As the tracking of clusters requires the implementation of complex recurrent algorithms and extensive use of temporary variables, this case study is intended to test the flexibility of the analysis framework of AViS. For this study, we performed a simulation of 10976 LJ atoms using FDPS \cite{iwasawa2016implementation}, and recorded a total of 10,000 snapshots over a simulation time of 21.5 ns. 

First, we identified the largest nucleated cluster in the final snapshot. The flexibility of DFP allows us to split up the algorithm into different generic functions, as shown in Fig. \ref{fig:nucwebs} (a). We used three different functions written in C++. Their contents are listed as follows:

\begin{figure}[ht]\centering
    \includegraphics[width=0.8\textwidth]{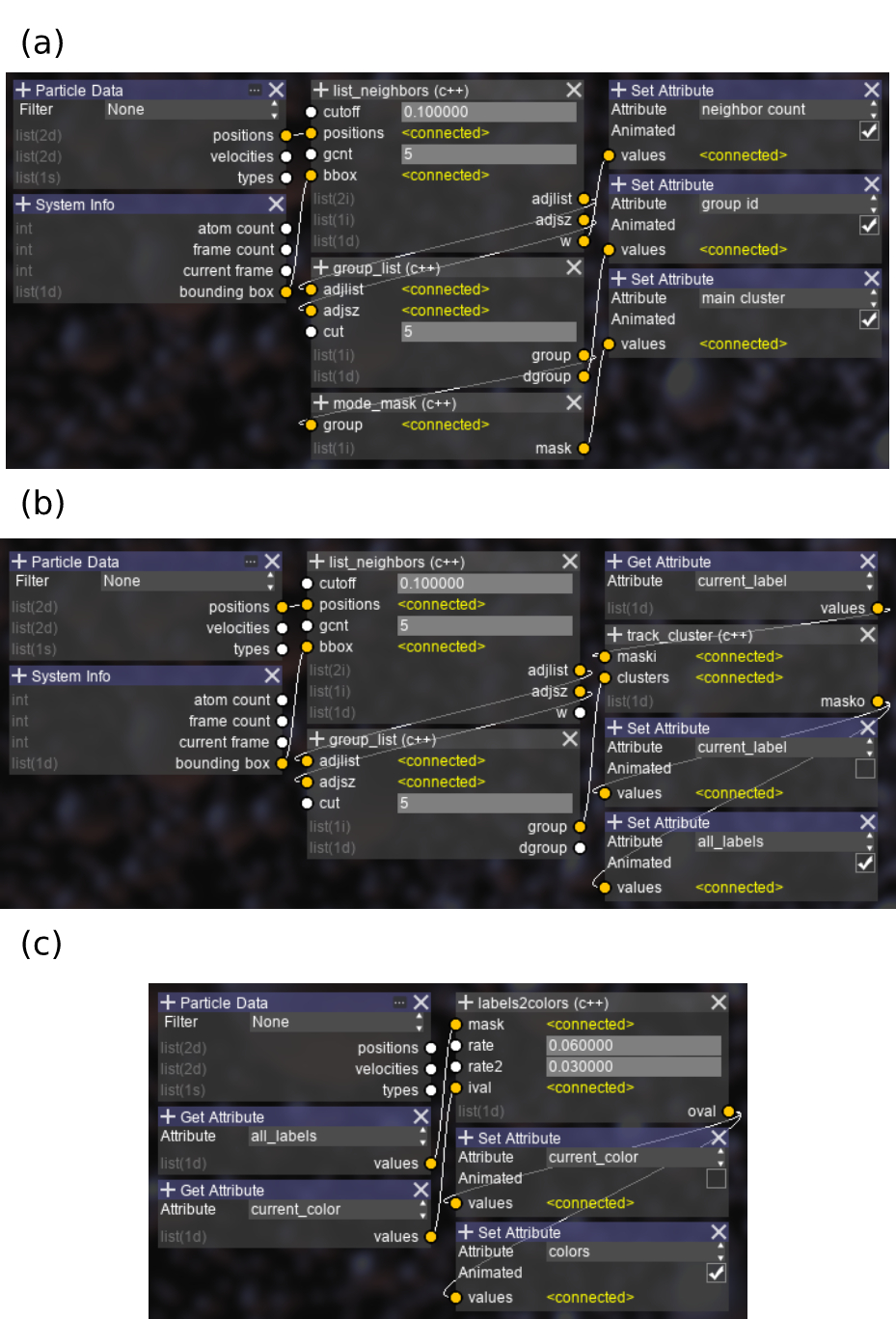}
    \caption{\label{fig:nucwebs} {\bf The analysis setup for LJ droplets.} (a) The largest cluster identification algorithm setup. (b) The cluster tracking algorithm setup. (c) The color graduation algorithm setup.}
\end{figure}

\begin{itemize}
    \item \textit{list\_neighbors}: Create a neighbor-list for each molecule with a user-defined cutoff.
    \item \textit{group\_list}: Apply a unique ID to each group of connected molecules.
    \item \textit{mode\_mask}: Label the molecules with the ID that appears most often.
\end{itemize}

\begin{figure}[ht]\centering
    \includegraphics[width=\textwidth]{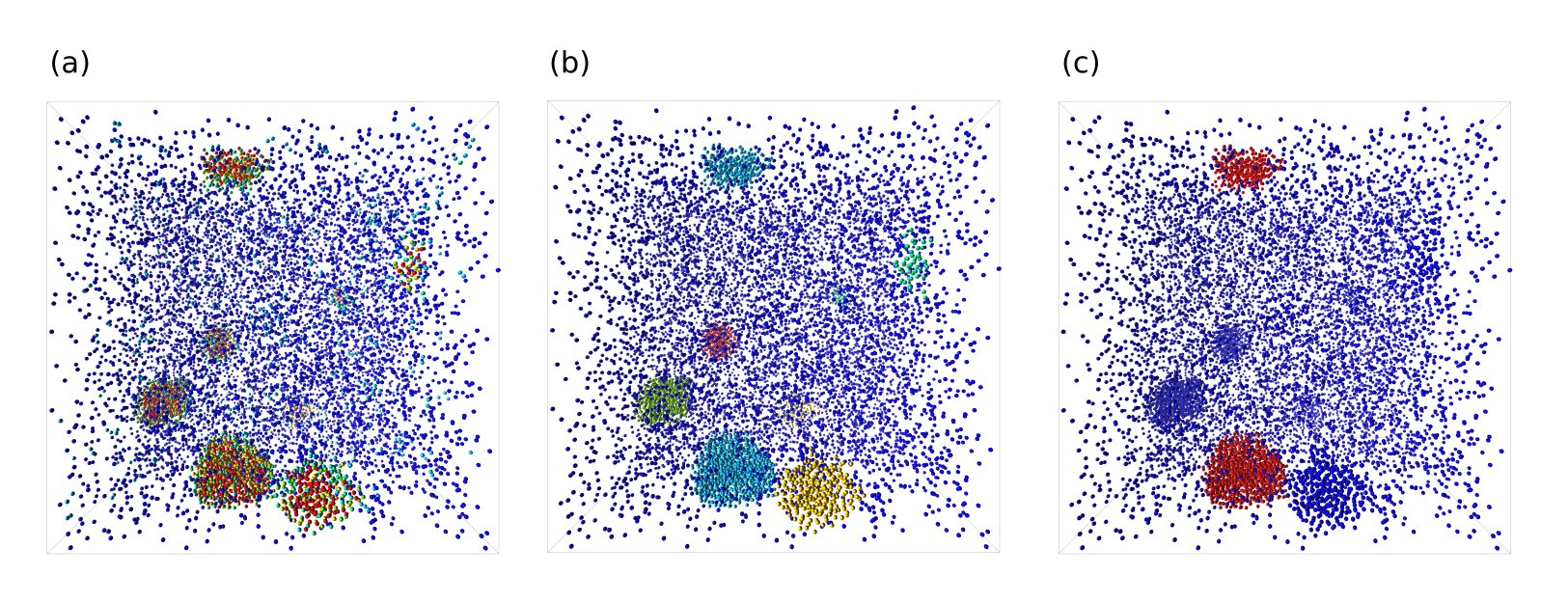}
    \caption{\label{fig:nuclast} {\bf Visualization of the attribute results of the largest cluster identification algorithm.} (a) The number of neighbors of each molecule. The color shifts from blue to red with an increasing number of neighbors. (b) The cluster index of each molecule. Each color corresponds to an unique cluster index. (c) The label mask for the largest cluster. The largest cluster is colored red. }
\end{figure}

The results of each step are saved as a different attribute for debugging, with their visualization results shown in Fig. \ref{fig:nuclast}. By splitting the algorithm into separate nodes, we are able to visually debug the results of each step. This is considerably easier than implementing the whole algorithm in one function.

Next, we tracked the formation of the identified cluster by devising a recurrent algorithm that updates the cluster label based on the next snapshot. The visual representation of the algorithm is shown in \nameref{s4_fig}. First, the cluster IDs for the current frame are computed. Next, the number of molecules in each cluster that is labeled as a cluster in the previous frame are counted. Finally, the cluster with the largest number of counted molecules is identified as the main cluster, and the labels are updated accordingly. If the count is 0, the cluster has dissipated, and so the labels are cleared. This algorithm is implemented in a new node:

\begin{itemize}
    \item \textit{track\_cluster}: Update the labels based on current cluster groups.
\end{itemize}

\begin{figure}[ht]\centering
    \includegraphics[width=\textwidth]{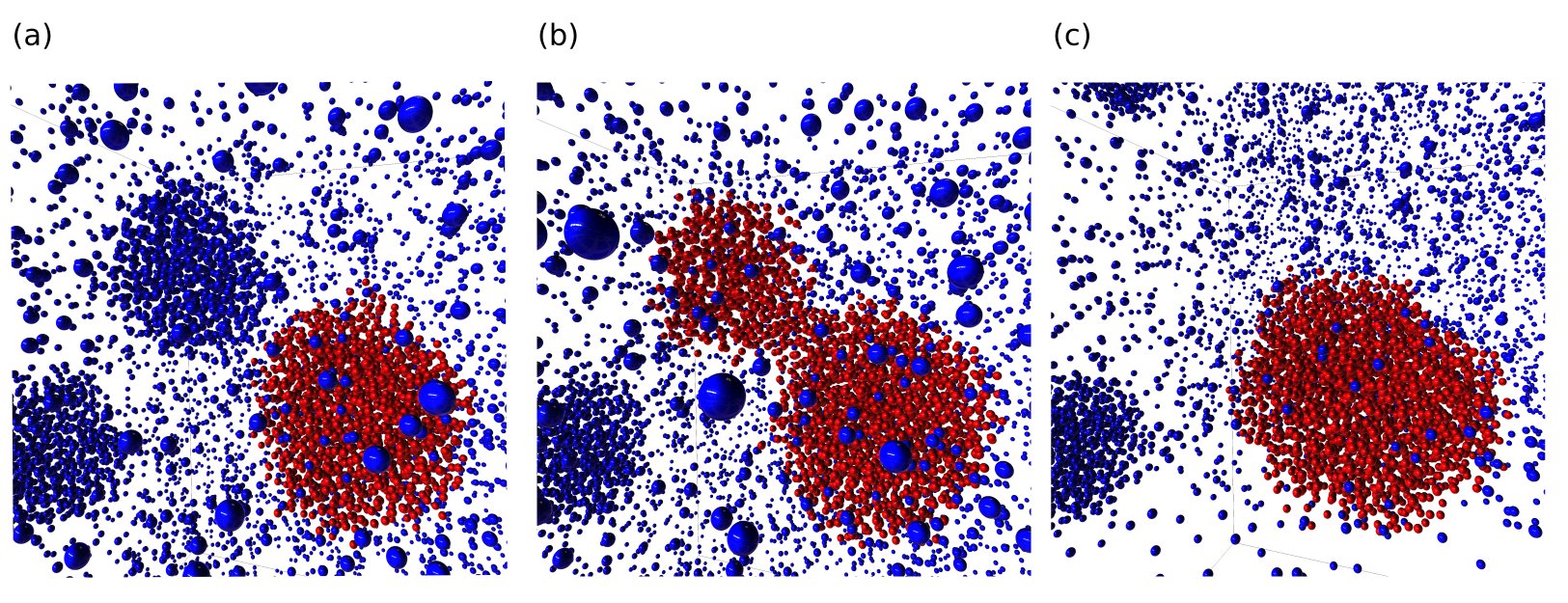}
    \caption{\label{fig:nucmask} {\bf Visualization of the results of the tracking algorithm.} In this algorithm, the main cluster is tracked backwards through time and colored red. (a) The final snapshot. The largest cluster is labeled. (b) The snapshot during the instant two clusters have combined. The clusters still share a common label. (c) The snapshot before the two clusters combined. The largest of the two is tracked and labeled.}
\end{figure}

As this algorithm still requires the grouping of clusters in each snapshot, we exploit the flexibility of the graph system by reusing the grouping functions in the previous step. The complete setup is shown in Fig. \ref{fig:nucwebs} (b). To make the algorithm recurrent, the same attribute (\textit{current\_label}) is set as both the input and output target of the \textit{track\_cluster} node, which is allowed in AViS. An additional attribute is also used to save the output for the whole trajectory, whose results are shown in Fig. \ref{fig:nucmask}.

Finally, we applied a gradient to each molecule based on its formation time, where atoms would fade to red as they join the cluster, and fade to blue when they leave. This is achieved by using a second recurrent function centered on a new node:

\begin{itemize}
    \item \textit{labels2colors}: Update the color of atoms based on their label.
\end{itemize}

\begin{figure}[ht]\centering
    \includegraphics[width=\textwidth]{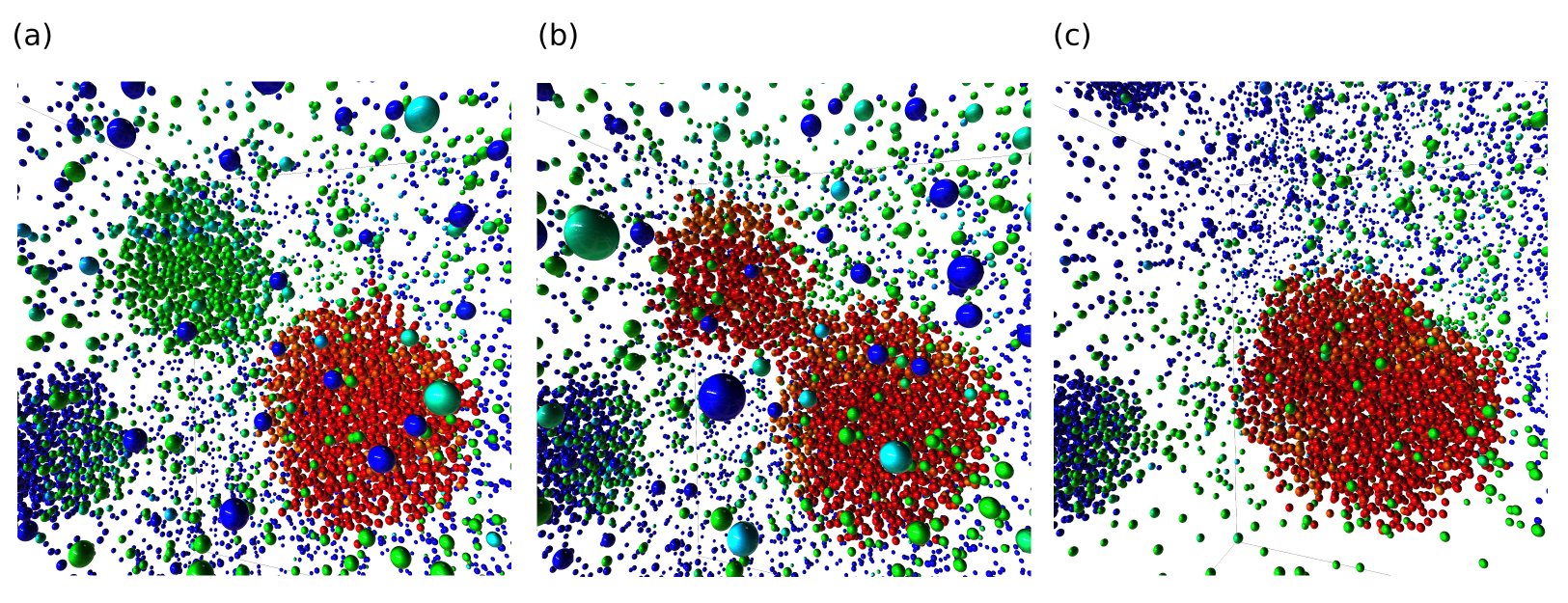}
    \caption{\label{fig:nucfade} {\bf The results of the color graduation algorithm.} A color gradient is applied so that atoms are colored blue to red based on how near they are to being a part of the main cluster. The same snapshots from Fig. \ref{fig:nucmask} is used. (a) A subcluster turns green right before merging with the tracked cluster (b). (c) A small green cloud surrounds the tracked cluster, indicating the constant flux of particles. }
\end{figure}

By running the setup forward through time, as shown in Fig. \ref{fig:nucwebs} (c), atoms fade into red or blue as they join or leave the cluster. However, we also require that atoms that are about to join a cluster fade from blue and vice versa. This is easily achieved in AViS by running the exact same algorithm backwards by setting the appropriate option. This shows the advantage of exposing the algorithm as nodes, in that it allows for changes in the execution flow without modifying the code. The attributes created from both loops are then combined, resulting in the output as shown in Fig. \ref{fig:nucfade}.

Other common processes in cluster analysis include finding the largest cluster at every point in time. As we already have all the components, we simply plug the \textit{group\_ids} into the \textit{mode\_mask} node. We can also offset the camera to point to the labeled cluster by calculating its center with a self-evident script and passing the result to the \textit{Set Camera Center} node.

To conclude, this case study showed that by using AViS' graph system, algorithms that involve loops over multiple modules that are otherwise difficult to achieve can be easily implemented. As the module implementations are completely isolated, each function can also be written with little to no reference to others, thus speeding up the analysis procedure.

\subsection*{\sffamily \large Visualization Flexibility}

The Mutually Coordinated Guest (MCG-1) algorithm for the study of the growth of methane clathrate hydrates was presented by Barnes \textit{et al.} \cite{barnes}. For effective analysis of the results, we would like to change the appearance of atoms based on the results, as well as create bonds forming the hydrate-like structures. Therefore, high interoperability between analysis and visualization modules is required. In this study, we used a portion of the same data created by the authors of the MCG-1 algorithm \cite{barnes2}, which contains a mixture of 2944 water atoms and 512 methane atoms, over 8000 frames.

\begin{figure}[ht]\centering
    \includegraphics[width=1\textwidth]{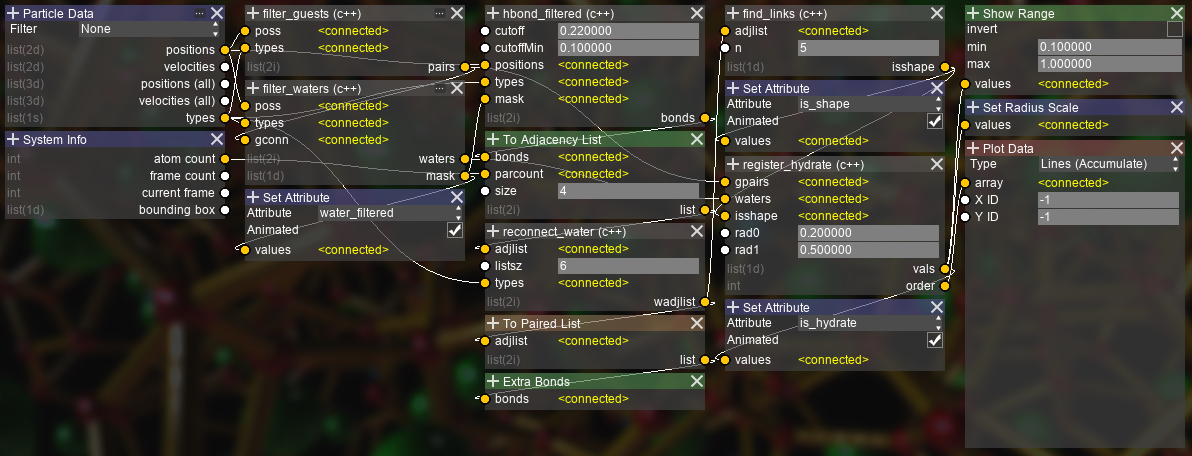}
    \caption{\label{fig:hydweb} {\bf The complete setup for the second case study.} The analysis setup for the hydrate formation case study, namely the MCG-1 algorithm.}
\end{figure}

The MCG algorithm looks for the structure shown in \nameref{s5_fig}. The algorithm is split into separate modules for ease of debugging, with each module explained below. The completed analysis graph is shown in Fig. \ref{fig:hydweb}.

Several nodes are created for this case study, one for each step of the algorithm:

\begin{itemize}
    \item \textit{filter\_guests}: Find pairs of guests within 9\AA  of each other.
    \item \textit{filter\_waters}: Find water molecules that are $45^{\circ}$ between each guest pair.
    \item \textit{hbonds\_filtered}: Find hydrogen bonds between selected water molecules.
    \item \textit{reconnect\_water}: Change O-H bonds to O-O bonds.  
    \item \textit{find\_links}: Look for bonds making up a \textit{n}-chain polygon. In this study, \textit{n} is set to 5.
    \item \textit{register\_hydrate}: Assign a unique label for caged guests and their oxygen atom structures.
\end{itemize}

Additionally, the following operational nodes provided in AViS are used for modifications on the visualization state:

\begin{itemize}
    \item \textit{Extra Bonds}: Create additional bonds between atoms.
    \item \textit{Show Range}: Hide atoms with an attribute value outside of a specified range.
    \item \textit{Set Radius Scale}: Modify the radius of atoms.
\end{itemize}

\begin{figure}[ht]\centering
    \includegraphics[width=\textwidth]{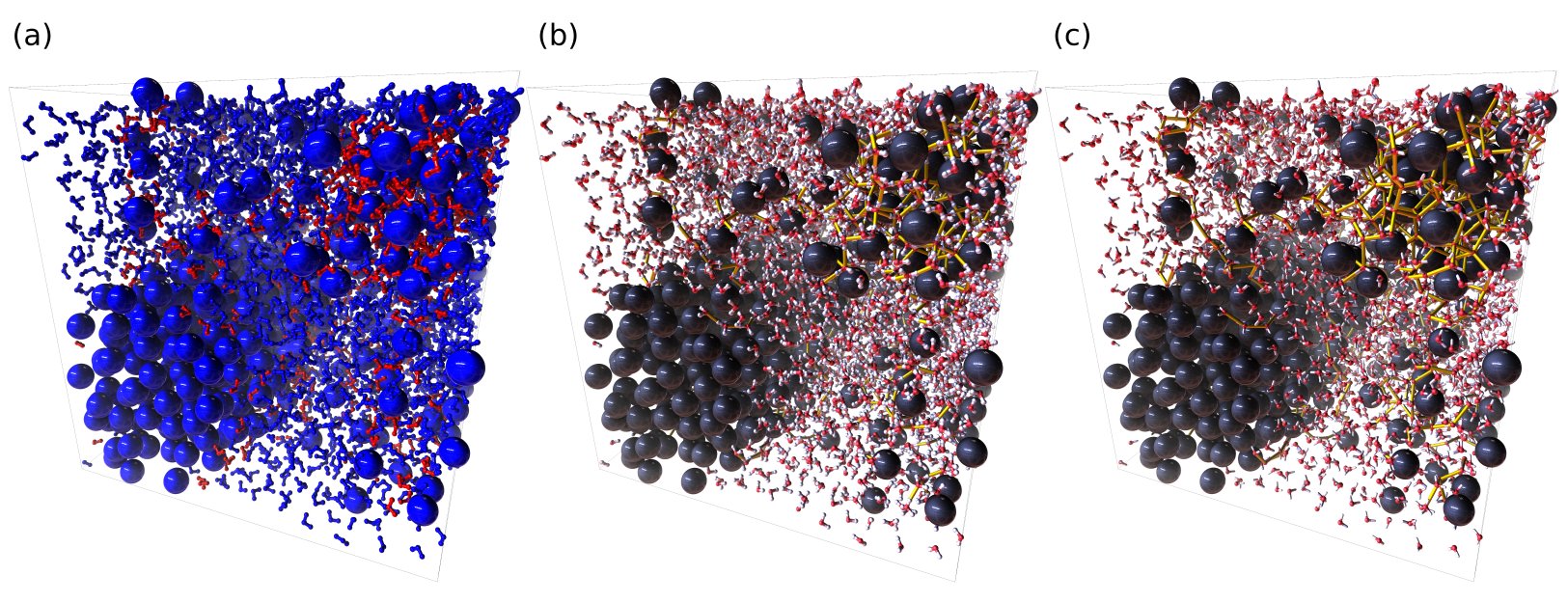}
    \caption{\label{fig:hydsteps} {\bf The intermediate results of the hydrate analysis.} Large spheres are methane molecules, while small connected spheres are O and H atoms of water molecules. (a) Water molecules identified by the MCG-1 algorithm are colored red. (b) Cage bonds are drawn in yellow in the form of O-H bonds. (c) Cage bonds are drawn in yellow in the form of O-O bonds. AViS is able to change between (b) and (c) by simply switching the connection of nodes. }
\end{figure}

As with the first case study, splitting the algorithm into parts allowed us to visually debug the validity of each step. As with the previous case study, all atom properties such as masks are saved as attributes, one of which can be seen in Fig. \ref{fig:hydsteps} (a). Also, the bonds created can be easily switched between O-H (Fig. \ref{fig:hydsteps} (b)) and O-O (Fig. \ref{fig:hydsteps} (c)) by simply changing the input source of the \textit{Extra Bonds} node. This further shows the advantage of exposing critical functions in the UI instead of as an API.

\begin{figure}[ht]\centering
    \includegraphics[width=\textwidth]{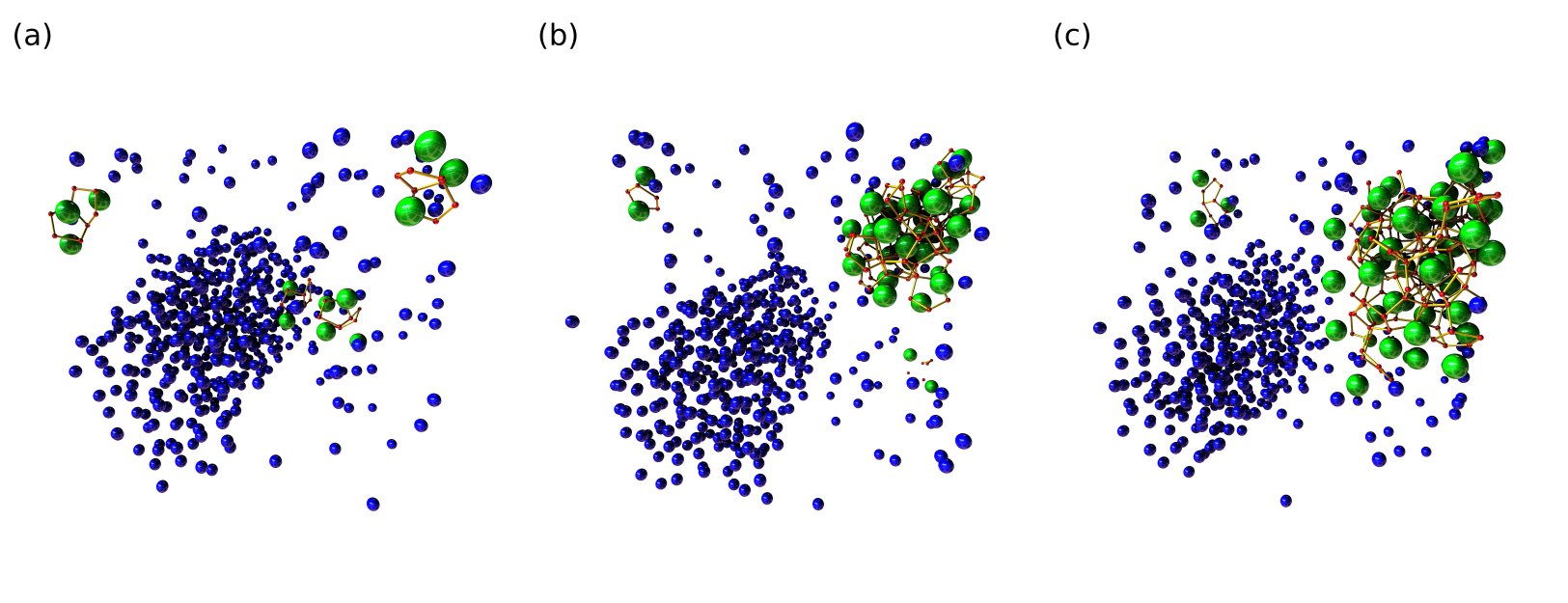}
    \caption{\label{fig:hydtrj} {\bf The progression of hydrate formation in the trajectory.} Unidentified and identified methane molecules are colored blue and green respectively, and unindentified water molecules are not visualized. Snapshots (a), (b), and (c), are taken at arbitrary frames throughout the trajectory. AViS caches the results of each analysis frame, so modifications including radius, color, and additional bonds can be animated.}
\end{figure}

By running the algorithm across the whole trajectory, the progression of the formation of hydrates can be visualized as shown in Fig. \ref{fig:hydtrj}. AViS automatically caches the result for each step, so when the calculation is finished, all results can be visualized in the window directly, including dynamic bonds, visibility, and size changes.

Additionally, a crucial step in hydrate nucleation analysis is identifying the formation time of the initial cluster by plotting the value of the order parameter through time. This can be achieved in AViS by using the \textit{Plot Data} node set to the \textit{Lines (Accumulate)} mode, taking the order parameter value as input. The output of the node is shown in \nameref{s6_fig}, which shows the point of hydrate formation at approximately frame 2000.

To conclude, this case study showed that by exposing essential functions in the UI, changes to the visualization state can be made instantly, with no need for recompilations or API lookups. As a result, visual debugging of algorithms can be performed easily, without extensive and potentially complex changes to the code.

\section*{Conclusions}

We have introduced Dataflow Programming for the analysis of MD data by creating AViS and demonstrated its flexibility in two complex case studies. We also showed that by abstracting the analysis scripts, we are able to substantially reduce the complexity of user code, while retaining the ability to interact with the visualization system. We believe that the DFP method has great potential, and while manual parallelization of single nodes using OpenMP is possible, we expect future work on parallelization of graph executions to further improve analysis performance. The source code as well as binaries of AViS for Windows, Linux, and MacOS, are hosted on Github at \url{https://github.com/avis-md/avis}.

\section*{Supporting information}

\paragraph*{S1 Table.}
\label{s1_table}
{\bf Examples of in-built nodes provided by AViS.} Custom nodes can be implemented by the user to interact with internal nodes, regardless of language.

\paragraph*{S1 Appendix.}
\label{s1_app}
{\bf a valid fortran analysis script that generates a sine wave.}

\paragraph*{S2 Appendix.}
\label{s2_app}
{\bf A valid Python analysis script that differentiates an array.}

\paragraph*{S3 Appendix.}
\label{s3_app}
{\bf A valid C++ analysis script that applies an exponential decay to a signal.}

\paragraph*{S1 Fig.}
\label{s1_fig}
{\bf Multi-language support for analysis. AViS handles data transfer between Fortran, Python, and C++ scripts automatically, including arrays as seen here.}

\paragraph*{S4 Appendix.}
\label{s4_app}
{\bf Comment syntax for Fortran scripts.}

\paragraph*{S5 Appendix.}
\label{s5_app}
{\bf Comment syntax for Python scripts.}

\paragraph*{S6 Appendix.}
\label{s6_app}
{\bf Comment syntax for C++ scripts.}

\paragraph*{S7 Appendix.}
\label{s7_app}
{\bf An example of a SSV file with rotational and potential attributes.} Each entry in the first line represents the type of data in each column.

\paragraph*{S2 Fig.}
\label{s2_fig}
{\bf Visualization and analysis of molecules with attribute data.} (a) By importing the orientation data as attributes with the Generic SSV format, liquid crystal molecules can be visualized without the need to write a custom plugin or importer. In this figure, a color gradient is also applied based on the \textit{roty} attribute. (b) Individual attributes can be used in analysis by utilizing the \textit{Get Attribute} node.

\paragraph*{S3 Fig.}
\label{s3_fig}
{\bf The rendering pipeline.} (a) Bounding quads are generated using a vertex shader. (b)(c) Surface information channels are generated using ray-tracing in a fragment shader. (d) A Physically-based Rendering (PBR) shader combines the channels into the final image. (e) By utilizing deferred shading, UI overlays and image effects can be added without re-drawing the whole scene, thus improving performance.

\paragraph*{S4 Fig.}
\label{s4_fig}
{\bf The visual representation of the tracking algorithm.} (a) The current frame. (b) The next frame. (c) The number of particles belonging to the previous cluster is counted. (d) The labels are updated for the cluster with the largest count.

\paragraph*{S5 Fig.}
\label{s5_fig}
{\bf MCG-1 algorithm representation.} A visual representation of the mutually coordinated guest order parameter used in the case study on clathrate hydrates. The algorithm is presented by Barnes \textit{et al.} \cite{barnes}).

\paragraph*{S6 Fig.}
\label{s6_fig}
{\bf The value of the order parameter across the whole trajectory.} By recording the number of classified molecules every step, we can identify the point of hydrate formation at about frame 2000.

\subsection*{Acknowledgements}

The authors would like to thank Y. Ono for generously providing us with the carbon nanotubes data for the creation of Fig. \ref{fig:ui}, as well as Dr. T. Nozawa for the Liquid Crystal data for the creation of \nameref{s2_fig}. Thanks are also given to Dr. M. Nakayama for his invaluable input on visualization techniques in OpenGL. We would also like to thank Prof. A. K. Sum for allowing us to use the data in the second case study on methane hydrates. Finally, we would like to thank all members of the Yasuoka Laboratory for trying out the software and giving invaluable suggestions and bug reports.



\end{document}